\documentstyle[epsf,preprint,pre,aps]{revtex}
\begin{document}
% uncomment next line for two-column draft mode
%\twocolumn[\hsize\textwidth\columnwidth\hsize\csname @twocolumnfalse\endcsname
\draft
\title{Cooperative Chiral Order in Copolymers of Chiral~and~Achiral~Units}
\author{J. V. Selinger}
\address{Center for Bio/Molecular Science and Engineering,
Naval Research Laboratory, Code 6900, \\
4555 Overlook Avenue, SW, Washington, DC  20375}
\author{R. L. B. Selinger}
\address{Department of Physics, Catholic University of America,
Washington, DC 20064}
\date{August 21, 1996}
\maketitle
\begin{abstract}
Polyisocyanates can be synthesized with chiral and achiral pendant groups
distributed randomly along the chains.  The overall chiral order,
measured by optical activity, is strongly cooperative and depends
sensitively on the concentration of chiral pendant groups.  To explain
this cooperative chiral order theoretically, we map the random copolymer
onto the one-dimensional random-field Ising model.  We show that the
optical activity as a function of composition is well-described by the
predictions of this theory.
\end{abstract}
\pacs{PACS numbers:  61.41.+e, 05.50.+q, 78.20.Ek, 82.90.+j}
% uncomment next line for two-column draft mode
%\vskip2pc]
\narrowtext

In a recent series of experiments, Green {\it et~al.} found that
polyisocyanates exhibit a very striking type of cooperative chiral
order~\cite{picreview}.  Polyisocyanates can be synthesized with any
combination of right-handed, left-handed, and achiral pendant groups
distributed randomly along the chains.  The overall chiral order of the
copolymers in solution, as measured by their optical activity, depends
sensitively on the concentrations of the different types of pendant
groups.  Polyisocyanates formed from a mixture of right- and left-handed
enantiomers follow a chiral ``majority rule''~\cite{majority}.  The
optical activity responds sharply to slight differences in the
concentrations of the enantiomers, and is dominated by whichever
enantiomer is in the majority.  Similarly, polyisocyanates with a mixture
of chiral (right-handed) and achiral pendant groups, known as
``sergeants-and-soldiers'' copolymers, are sensitive to very low
concentrations of chiral pendant groups~\cite{ssexp}.  They have a
substantial optical activity even when the concentration of chiral units
is less than 1\%.  The optical activity of both
systems shows a high degree of cooperativity within the copolymers.

In an earlier paper, we proposed a theory for the majority-rule 
system~\cite{theory1}.  The basis of our theory was a mapping of the 
random copolymer onto the random-field Ising model, a standard model in 
the theory of random magnetic systems~\cite{rfim1,rfim2}.  Using this 
model, we predicted the chiral order of the copolymer as a function of 
enantiomer concentration, in quantitative agreement with the experiments, 
and showed that the sharpness of the majority-rule curve is determined by 
two energy scales associated with the chiral packing of monomers.  In 
this paper, we show that the same type of theory can also explain the 
chiral order of the sergeants-and-soldiers system.  Using the 
random-field Ising model, we predict the chiral order of these copolymers 
as a function of the concentration of chiral pendant groups.  We show 
that the measured optical activity is well-described by this prediction 
with a reasonable choice of parameters.

As the material properties of polymers generally depend on a wide
variety of chemical and geometric parameters, the present result is
remarkable in that it explains the complex behavior of polyisocyanates
in terms of a one-dimensional theory with only two
material-dependent parameters.
That such a simple theory describes not only
one but two different classes of polyisocyanates indicates that
the theory accurately represents the physics of cooperative chiral order.
Insight from the theory can guide the optimization of material
properties for technological applications, as discussed below.

\begin{figure}
\centering\leavevmode\epsfbox{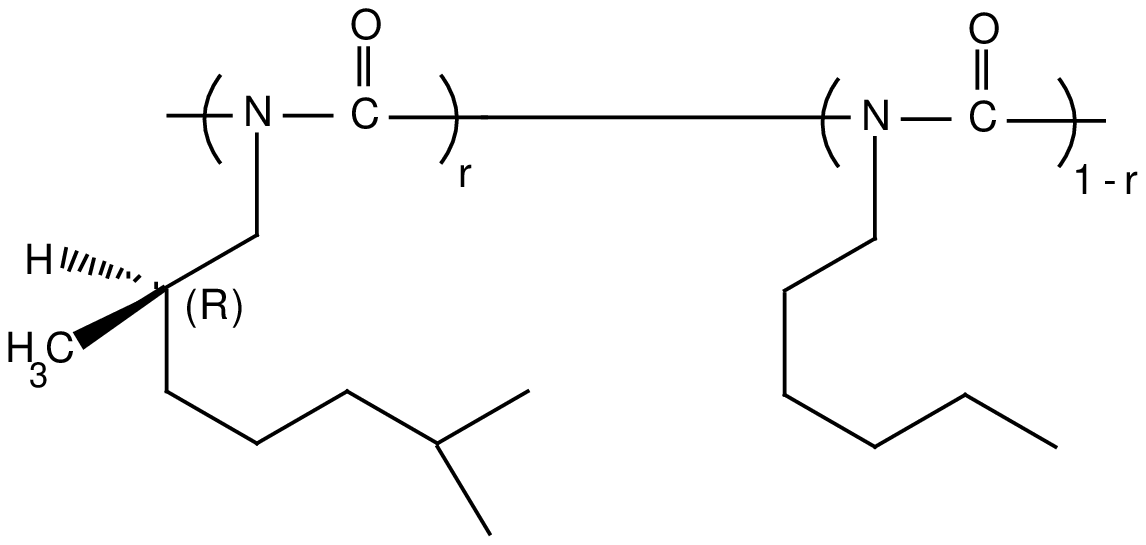}
\caption{Molecular structure of the sergeants-and-soldiers copolymer that
was studied in Ref.~\protect\cite{ssexp}.  A fraction $r$ of the pendant
groups are chiral (right-handed), and $1-r$ are achiral.}
\label{fig1}
\end{figure}

\begin{figure}
\epsfbox{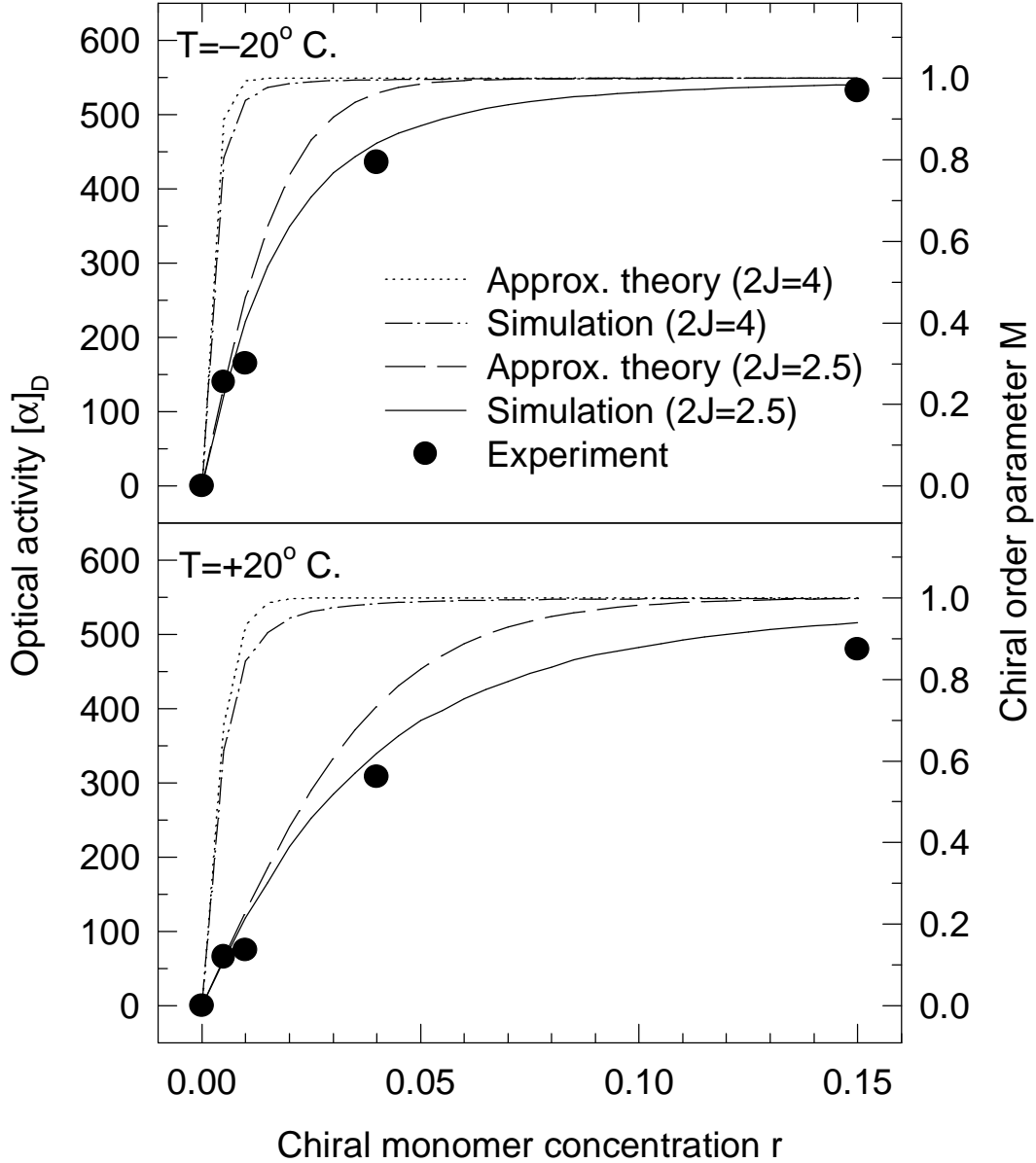}
\caption{The symbols show the optical activity $[\alpha]_D$ of the 
sergeants-and-soldiers copolymer as a function of the chiral monomer 
concentration $r$ for $T=\pm20^\circ$~C. from Table~I of 
Ref.~\protect\cite{ssexp}.  (Data for $r=0.373$ and $r=1$ are not shown.)
The first pair of lines (dotted and dot-dashed) shows the predictions for
the chiral order parameter $M$ from the approximate theory and the
simulation with $2h=0.4$~kcal/mol, $2J=4$~kcal/mol, and $N=1000$.
The second pair of lines (dashed and solid) shows the predictions for
$M$ from the approximate theory and the simulation with $2h=0.4$~kcal/mol,
$2J=2.5$~kcal/mol, and $N=1000$.  The relative scale of the two vertical
axes is the maximum optical activity observed.  The second choice of
parameters gives good agreement between theory and experiment.}
\label{fig2}
\end{figure}

The sergeants-and-soldiers copolymer studied by Green {\it et~al.} is
shown in Fig.~\ref{fig1}~\cite{ssexp}.  This polymer consists of a
carbon-nitrogen backbone with a pendant group attached to each monomer.
The pendant groups can be either chiral or achiral; the backbone itself
is achiral.  Steric constraints force the polymer into a helical
conformation, which can be either right- or left-handed.  The helical
structure of the polymers can be investigated experimentally by measuring
their optical activity in solution.
Because the right- and left-handed helices
rotate the polarization of light in opposite directions, the optical
activity is proportional to the difference in the proportion of right-
and left-handed helices.  If all the pendant groups are achiral, the
right- and left-handed helices have the same energy.  A long chain then
consists of domains of right- and left-handed helicity, separated by
occasional helix reversals.  On average, the proportions of right- and
left-handed helices are equal, leading to zero net optical activity.
However, if a fraction $r$ of the pendant groups are chiral, there is a
preference for one sense of helicity, which leads to a net optical
activity.  Green {\it et~al.} measured the optical activity as a
function of $r$.  Their results are plotted in Fig.~\ref{fig2}.  The
measured optical activity is extremely sensitive to low concentrations of
chiral pendant groups:  $r=0.15$ gives almost the same optical activity
as the pure chiral homopolymer with $r=1$ (not shown in Fig.~\ref{fig2}),
and even $r=0.005$ gives a substantial optical activity.

To explain this cooperative chiral order, we map the random copolymer
onto the one-dimensional random-field Ising model.  In this mapping, the
Ising spin $\sigma_i$ corresponds to the local sense of the polymer
helix at monomer $i$:  $\sigma_i=+1$ represents a right-handed helix and
$\sigma_i=-1$ a left-handed helix.  The Hamiltonian for a copolymer of
length $N$ can be written as
\begin{equation}
H=-J\sum_{i=1}^{N-1}\sigma_i\sigma_{i+1}-\sum_{i=1}^{N}h_i\sigma_i.
\end{equation}
The first term in this Hamiltonian gives the energy cost of a helix
reversal, and the second term gives the local chiral bias, an effective
field favoring one sense of the helix.  If monomer $i$ is chiral (with
probability $r$) then its chiral bias is $h_i=+h$, and if the monomer is
achiral (with probability $1-r$) then $h_i=0$.  The field $h_i$ is a
quenched random variable; it is fixed by the polymerization of each chain
and does not change in response to changes in $\sigma_i$.  The
magnetization of the Ising model,
\begin{equation}
M=\frac{1}{N}\left\langle\sum_{i=1}^{N}\sigma_i\right\rangle,
\end{equation}
corresponds to the chiral order parameter that is measured by the optical
activity.  To predict the optical activity, we must calculate $M$.

Before proceeding with the calculation, we note that
$h$ and $J$ are approximately known from earlier
studies.  The parameter $2h$ is the energy cost of a
right-handed monomer in a left-handed helix.  Through molecular modeling,
Lifson {\it et~al.} calculated $2h\approx0.4$~kcal/mol~\cite{molecmodel}.
The parameter $2J$ is the energy cost of a helix reversal.  Through fits
of the optical activity of deuterated homopolymers as a function of
temperature $T$, Lifson {\it et~al.} obtained
$2J\approx4$~kcal/mol~\cite{deuterium}.  The precise value of $2J$
depends on the solvent:  it is higher for hexane and lower for
chloroform.  The majority-rule experiment used
hexane~\cite{majority}, while the sergeants-and-soldiers experiment
used chloroform~\cite{ssexp}.  The solvent dependence will be
discussed further below.  At room temperature, $2J$ is much greater than
$k_B T\approx0.6$~kcal/mol, but $2h<k_B T$.  Thus,
helix reversals are quite costly in energy, while the chiral bias of a
single monomer is fairly small.

To calculate the chiral order parameter $M$, we follow a procedure
analogous to our earlier calculation for the majority-rule
system~\cite{theory1}. First, we note that each chain consists of domains
of uniform helicity $\sigma_i$.  Suppose that each domain has length $L$,
which is to be determined.  Each domain responds to the total chiral
field $h_{\rm tot}=\sum h_i$ of the monomers in it.  Because the domain
is uniform, the response is $M(h_{\rm tot})=\tanh(h_{\rm tot}/k_B T)$,
equivalent to a single spin in a magnetic field.  Averaging over the
probability distribution $P(h_{\rm tot})$, we obtain
\begin{equation}
M=\int_{-\infty}^{\infty}dh_{\rm tot}P(h_{\rm tot})
\tanh\left(\frac{h_{\rm tot}}{k_B T}\right).
\label{integral}
\end{equation}
The probability distribution $P(h_{\rm tot})$ is a binomial distribution.
It can be approximated as follows:

\paragraph*{Case A:  $rL\ll 1$.}  In this case, most domains do not have
any chiral monomers.  We have $h_{\rm tot}=0$ with probability $(1-r)^L$,
$h_{\rm tot}=h$ with probability $Lr(1-r)^{L-1}$, and
$h_{\rm tot}\geq 2h$ with negligible probability.  We can therefore
expand Eq.~(\ref{integral}) in powers of $rL$ to obtain
\begin{equation}
M\approx rL\tanh\frac{h}{k_B T}.
\label{casea}
\end{equation}

\paragraph*{Case B:  $rL\gg 1$.}  In this case, most domains have many
chiral monomers.  The binomial distribution $P(h_{\rm tot})$ can then be
approximated by a Gaussian with mean $hLr$ and standard deviation
$h[Lr(1-r)]^{1/2}$.  Around the peak of the Gaussian, variations in the
tanh in Eq.~(\ref{integral}) are negligible.  Hence, we obtain
\begin{equation}
M\approx\tanh\frac{rLh}{k_B T}.
\label{caseb}
\end{equation}

Note that Eqs.~(\ref{casea}) and~(\ref{caseb}) resulting from cases~A
and~B are quite similar.  The limit of Eq.~(\ref{caseb}) for small $rL$
is $M\approx rLh/k_B T$, which is equivalent to Eq.~(\ref{casea})
provided that $h\lesssim k_B T$.  This is valid in the experiments.
Hence, it is a good approximation to use Eq.~(\ref{caseb}) for the whole
range of $rL$.

We must now estimate the characteristic domain size $L$.  The domain size
is determined by the density $1/L$ of domain boundaries.  Two mechanisms
contribute to $1/L$:  (a) the density $1/L_{\rm th}$ of helix reversals
induced by thermal fluctuations and (b) the density $1/N$ of chain ends.
For low densities, these mechanisms should be additive (although they
will interact for higher densities).  Hence,
\begin{equation}
\frac{1}{L}\approx\frac{1}{L_{\rm th}}+\frac{1}{N}.
\label{domainsize}
\end{equation}
For $r=0$, the thermal domain size is $L_{\rm th}=e^{2J/k_B T}$.  For
small nonzero $r$, we can use this value of $L_{\rm th}$ as an
approximation.  If $2J=4$~kcal/mol, this size is approximately 2800
monomers at $T=-20^\circ$~C. and 960 monomers at $T=+20^\circ$~C.  By
comparison, the chain length $N$ ranges from 3000 to 10000 in the
experiment~\cite{ssexp}.  For these values of $L_{\rm th}$ and $N$, the
domain size $L$ is dominated by $L_{\rm th}$ but is somewhat sensitive to
$N$.

We can point out an important difference between the
sergeants-and-soldiers system and the majority-rule system considered in
our earlier paper~\cite{theory1}.  In the majority-rule system, the
copolymer has both right-handed monomers with $h_i=+h$ and left-handed
monomers with $h_i=-h$.  The competition between these fields of opposite
signs gives a random-field domain size
$L_{\rm rf}\approx(J/h)^2\approx 100$.  Because $L_{\rm rf}$ is much less
than $L_{\rm th}$ and $N$, the domain size is limited by random-field
effects.  By contrast, in the sergeants-and-soldiers system, the
copolymer has only one type of chiral monomer, so there is no competition
between fields of opposite signs.  For that reason, the
sergeants-and-soldiers copolymer has no random-field domain size; its
domain size is limited only by thermal fluctuations and by the chain
length.  The domain size is therefore much larger in the
sergeants-and-soldiers system than in the majority-rule system.

Equations~(\ref{caseb}) and~(\ref{domainsize}) give a
prediction for $M$ as a function of $r$, $h$, $J$, and $N$. This 
prediction involves certain approximations---in particular, it neglects 
the dependence of $L$ on $r$ and $h$.  To test this approximate 
prediction, we perform numerical simulations of the 
random-field Ising model.  In the simulations, we construct an 
explicit realization of the random field $h_i$, then calculate the 
partition function and order parameter using transfer-matrix 
techniques.  We then average the order parameter over at least 1000 
realizations of the random field.  In the first series of simulations, we 
use $2h=0.4$~kcal/mol and $2J=4$~kcal/mol, as in our 
earlier simulations of the majority-rule system.  As discussed above, 
these parameters are expected to be approximately correct for the 
sergeants-and-soldiers system.  Figure~\ref{fig2} shows the simulation
results together with the predictions of the approximate theory for
$N=1000$, $T=\pm20^\circ$~C., and $r=0$ to $0.15$.  In most respects,
the approximate theory agrees with the simulation.  In particular, the
approximate theory predicts the slope $dM/dr$ at $r=0$ in
agreement with the simulation.  (Results for $N=100$, not shown, give
equally good agreement between the approximate theory and the simulation.)
There is some discrepancy between the approximate theory and 
the simulation for larger values of $r$.  This discrepancy
arises because the simulation implicitly includes the dependence of $L$
on $r$ and $h$, so it is more precise than the approximate theory.

Because we have fairly good agreement between the approximate theory and
the simulation, we can now compare the theory and simulation with the
experimental data in Fig.~\ref{fig2}.  For this comparison, the relative
scale of the optical-activity axis and the order-parameter axis is the
maximum optical activity observed.  With the parameters
$2h=0.4$~kcal/mol and $2J=4$~kcal/mol, the theory and simulation do not
agree with the experiment.  The prediction for the chiral
order parameter saturates much more rapidly than the
data for the optical activity.
If we used the experimental chain length $N=3000$ to $10000$
instead of $1000$, the disagreement would be
even slightly worse.  There may be two reasons why this choice of
parameters gave good agreement between theory and experiment in our
earlier study of the majority-rule system but poor agreement for the
sergeants-and-soldiers system.  First, the chiral order of the
sergeants-and-soldiers system is dominated by the thermal domain size
$L_{\rm th}=e^{2J/k_B T}$, which is exponentially sensitive to $J$.
Thus, small errors in $J$ have a large effect on the theoretical
prediction.  By comparison, the chiral order of the
majority-rule system is dominated by the random-field domain size
$L_{\rm rf}\approx(J/h)^2$, which is less sensitive to small errors in
$J$.  Second, as noted above, the two experiments were done in different
solvents:  the majority-rule experiment used hexane while the
sergeants-and-soldiers experiment used chloroform.  Experimental studies
of deuterated homopolymers show that the helix reversal energy is
lower in chloroform than in hexane~\cite{deuterium}.  Hence, the
disagreement between theory and experiment is probably due to errors in
the parameters, particularly $J$.

\begin{figure}
\epsfbox{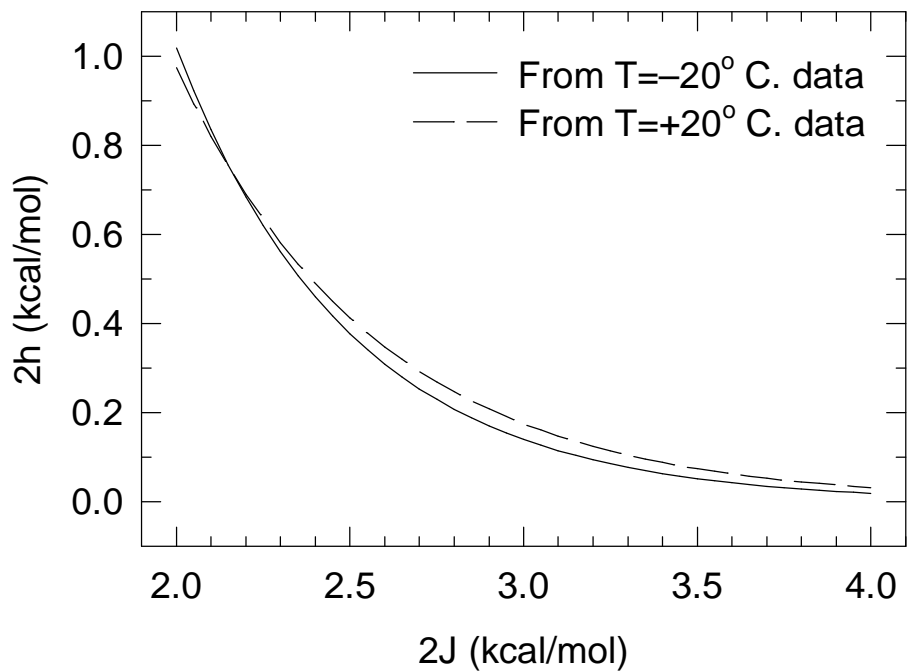}
\caption{Loci of values of $2h$ and $2J$ that give good 
agreement between theory and experiment at $T=\pm20^\circ$~C.}
\label{fig3}
\end{figure}

To find more accurate values of $h$ and $J$, 
we fit the approximate theory to the experimental data using the 
following procedure.  First, we normalize the data for the optical 
activity by the maximum optical activity observed, to obtain an estimate 
of $M$.  Next, we determine $dM/dr$ at $r=0$ from the data at each 
temperature.  This gives $dM/dr\approx 54$ at $T=-20^\circ$~C. and 
$dM/dr\approx 26$ at $T=+20^\circ$~C.  The approximate theory gives 
$dM/dr=hL/k_B T=(h/k_B T)e^{2J/k_B T}$ in the limit of large $N$.  Thus, 
by setting the theoretical prediction for $dM/dr$ equal to the 
experimental value, we obtain a locus of allowed points in the $(h,J)$ 
plane for each temperature.  Those loci are plotted in Fig.~\ref{fig3}.  
Note that those two curves lie almost on top of each other.  They 
intersect at $2h=0.7$~kcal/mol and $2J=2.2$~kcal/mol, but that 
intersection is very ill-defined numerically.  Any point along those 
curves would give a good fit to the experimental data.  For example, 
$2h=0.4$~kcal/mol and $2J=2.5$~kcal/mol is a reasonable choice, which 
agrees with the molecular modeling~\cite{molecmodel} and is roughly 
consistent with the experiments on deuterated 
homopolymers~\cite{deuterium}.  Other points on the curves with lower 
values of $2h$ and higher values of $2J$ would also be satisfactory.

To test the agreement between theory and experiment for this choice of
parameters, we perform a series of simulations with
$2h=0.4$~kcal/mol, $2J=2.5$~kcal/mol, $N=1000$, $T=\pm20^\circ$~C., and
$r=0$ to $0.15$.  (With these parameters, the thermal domain
size is $L_{\rm th}=144$ at $T=-20^\circ$~C. and $L_{\rm th}=73$ at
$T=+20^\circ$~C., which is much less than $N$, so the results are not
sensitive to $N$.)  The approximate theory and the simulation results are
plotted together with the experimental data in Fig.~\ref{fig2}.
With these parameters, the approximate theory and the simulation both
agree with the experiment.  As expected, for larger values of
$r$, the simulation agrees with the experiment even better than the
approximate theory does.  Thus, the measured optical activity is
well-described by our theoretical approach with reasonable parameters.

This theory of cooperative chiral order may be useful for technological 
applications.  Zhang and Schuster have proposed a mechanism for an 
optical switch, which would be controlled by circularly
polarized light~\cite{switch}.  The majority-rule system of 
polyisocyanates would be well-suited for this application
because of the sharp dependence of optical activity on composition
near the racemic point.  To optimize polyisocyanates for this 
application, one must know $h$ and $J$ for a variety of pendant 
groups.  So far, $h$ and $J$ have only been measured in experiments on 
deuterated homopolymers, in which $h\lesssim k_B T/L$~\cite{deuterium}.  
If $h$ is larger, the optical activity of a homopolymer is saturated
at $M=1$, and no information can be extracted from it.  Hence, larger 
values of $h$ can only be calculated by molecular 
modeling~\cite{molecmodel}.  Our theory of sergeants-and-soldiers 
copolymers provides a new approach for determining $h$ and $J$.  Future 
experiments can synthesize sergeants-and-soldiers copolymers with two 
chain lengths, $N\ll L_{\rm th}$ and $N\gg L_{\rm th}$, and several 
values of $r$.  The parameter $h$ can be determined by fitting the 
optical activity of the short chains, and the 
combination $(h/k_B T)e^{2J/k_B T}$ can be determined by fitting the 
optical activity of the long chains.  Through this procedure, experiments 
can characterize $h$ and $J$ of a variety of polyisocyanates for use in 
an optical switch.

In conclusion, we have shown that our theoretical approach, based on the
random-field Ising model, can explain cooperative chiral order in the
sergeants-and-soldiers system as well as the majority-rule system.  The
sharp increase of the optical activity as a function of chiral monomer
concentration depends on the characteristic domain size $L$, which is
determined both by thermal fluctuations and by the chain length.  This
increase is well-described by our theoretical prediction with suitable
values of the energy scales $h$ and $J$.

We thank M.~M.~Green, B.~R.~Ratna, J.~M.~Schnur, R.~Shashidhar, and
M.~S.~Spector for many helpful discussions.  This research was supported
by the Naval Research Laboratory and the Office of Naval Research.

\end{document}